\documentclass[twocolumn,amsmath,amssymb,prd,nofootinbib,10pt,superscriptaddress]{revtex4-1}
\usepackage{ graphicx, float,amsmath, amsmath,  amsfonts, amssymb, xcolor, dsfont, braket}
\usepackage[export]{adjustbox}
\usepackage{comment}

\def\be{\begin{equation}}
\def\ee{\end{equation}}
\def\bea{\begin{eqnarray}}
\def\eea{\end{eqnarray}}
\def\bse{\begin{subequations}}
\def\ese{\end{subequations}}

\usepackage[breaklinks, colorlinks,linkcolor=black, citecolor=black,urlcolor = black]{hyperref}
\usepackage[labelsep=period]{caption}
\usepackage[normalem]{ulem}

\usepackage{amsmath}

\begin{document}

{\small{\textsc{Matches published version in Phys. Rev. D.}}}

\title{Squeezing of light from Planck-scale physics}

\author{Danilo Artigas}%
\email{danilo.artigas-guimarey@universite-paris-saclay.fr}
\affiliation{%
Universit\'e Paris-Saclay, CNRS, Institut d'Astrophysique Spatiale, 91405, Orsay, France
}%
\affiliation{%
Institute of Theoretical Physics, Jagiellonian University, {\L}ojasiewicza 11, 30-348 Cracow, Poland
}%

\author{Killian Martineau}%
 \email{martineau@lpsc.in2p3.fr}
\affiliation{%
Laboratoire de Physique Subatomique et de Cosmologie, Universit\'e Grenoble-Alpes, CNRS-IN2P3\\
53, avenue des Martyrs, 38026 Grenoble, France
}%

\author{Jakub Mielczarek}%
\email{jakub.mielczarek@uj.edu.pl}
\affiliation{%
Institute of Theoretical Physics, Jagiellonian University, {\L}ojasiewicza 11, 30-348 Cracow, Poland
}%


\begin{abstract} 
In this article, the possibility of generating nonclassical light due to Planck-scale effects is considered. For 
this purpose, a widely studied model of deformation of the Heisenberg uncertainty relation is applied to single-mode and multimode lights. The model leads to a deformed dispersion relation, which manifests in an advancement in the time of arrival of photons. The key finding is that the model also leads to an oscillatory pattern of squeezing of the state of light. Furthermore, while the amplitude of the oscillations is constant for energy eigenstates, it exhibits linear growth over time for coherent states with the annihilation operator eigenvalue $\alpha \neq0 $. This second case leads to the accumulation of squeezing and phase-space displacement, which can be significant for astrophysical photons. In particular, for $\alpha \sim 1$, coherent light in the optical spectrum emitted at megaparsec 
distances would acquire squeezing with the amplitude of the order unity. This suggests that measurements of the nonclassical properties of light originating from distant 
astrophysical sources may open a window to test these predictions. 
\end{abstract}
\maketitle

\section{Introduction}

The concept of a minimal length is one of the most widely 
examined potential manifestations of quantum characteristics 
of spacetime. The premise here is that there exists a finite 
limit to the precision at which space can be probed,
anticipated to be in the vicinity of the Planck length, 
$l_{\text{Pl}}\approx 1.62 \cdot 10^{-35}$ m. This idea originates 
from quantum gravity considerations, which suggest that at such 
extreme scales, traditional concepts of spacetime break down 
and give way to quantum effects.

The idea of a minimal length has been successfully integrated 
into a self-consistent special-relativistic framework, commonly 
known as doubly special relativity \cite{Magueijo:2001cr,
Amelino-Camelia:2002cqb}. This paradigm introduces 
a new invariant scale in addition to the speed of light, typically 
the Planck length, to reconcile quantum mechanics with relativity. 
In the broader landscape of quantum gravity theories, some approaches,
such as loop quantum gravity (LQG), propose a possible generalization 
of this concept to a general-relativistic context, thus extending its 
scope and potential implications. Nevertheless, the actual behavior 
and impact of this minimal length scale in the dynamical sector 
remains a subject of ongoing study. 

The concept of maximal spatial resolution is naturally implementable 
in the realm of quantum mechanics, particularly in relation to the 
Heisenberg uncertainty principle. This principle dictates inherent 
limits to the precision with which pairs of canonical variables, 
such as generalized position and momentum, can be simultaneously measured. 
By employing a generalized uncertainty principle (GUP) \cite{Bojowald:2011jd,
Bosso:2023aht}, it is feasible to integrate a minimal length scale into conventional 
quantum mechanics, thereby suggesting a fundamental limit to our ability 
to precisely measure position. Nonetheless, this adaptation inevitably 
implies that the commutation relation between the canonical variables, 
such as the generalized position and momentum, will undergo alterations. 

Modifications to the commutation relation are understood to correlate 
with the nonlinear geometry of the corresponding phase space. Notably, 
the existence of a minimum length, a potential hallmark of quantum gravity 
theories, is intimately associated with a curvature in the momentum component 
of the phase space. This relationship provides a geometric interpretation 
for the minimal quantum length \cite{Wagner:2021bqz}.

This concept is not confined to the domain of point particles -- it extends 
into the broader field theoretical context. An instance of this expansion 
can be found in the LQG-inspired polymer quantization. In this scenario, 
studies such as those presented in Refs. \cite{Hossain:2010wy,Hossain:2010eb} 
analyzed cylindrical (i.e., $\mathbb{R}\times \mathbb{S}$) deformations of 
the scalar field phase space. Other works in the context of string theory and quantum gravity also generalize the notion of phase space to deformed (possibly curved) ones \cite{Freidel:2013zga,Freidel:2017xsi,Riello:2017iti,Dittrich:2017nmq,Haggard:2015ima,Bonzom:2014wva,Dupuis:2019yds}. The nonlinear field space theory 
(NFST) research program, as outlined in Ref. \cite{Mielczarek:2016rax}, is 
committed to methodically explore the implications of fields having a 
nontrivial phase space. In the context of NFST, the case of $\mathbb{S}^2$ 
phase-space generalization of the scalar field theory has been primarily 
considered \cite{Artigas:2020hfj,Artigas:2020fab}. 

The effects of a GUP can lead to phenomenological consequences, currently 
under intense investigation in the context of multimessenger astronomy 
\cite{Addazi:2021xuf}. The physical consequences that are primarily analyzed 
are deformed dispersion relations and vacuum birefringence.

Here, we focus our attention on the properties of quantum states 
of light. Specifically, we focus on single and multimode light 
with GUP in a field theoretical context. Potential manifestations 
of the GUP have been previously explored in the particle context 
in Ref. \cite{Pikovski:2011zk}. Our aim is to build upon these existing 
insights and delve deeper into the understanding of light under 
the influence of GUP.

This article is structured as follows: In Sec. \ref{Sec:model}, we 
introduce the model of the Generalized Uncertainty Principle (GUP) and 
the ensuing single-mode Hamiltonian. Following this, we carry out a 
perturbative analysis of the time evolution of a single-mode light 
in Sec. \ref{Sec:evolution}. Based on these findings, we then examine 
the quantum squeezing of various states due to the GUP in Sec. 
\ref{Sec:squeezing}. As a supplement to the squeezing effect, Sec. 
\ref{Sec:dispersion} delves into how the GUP modifies the dispersion 
relation for a multimode light state. In Sec. \ref{Sec:experiments}, 
we consider the potential experimental implications of our derived predictions. 
Last, in Sec. \ref{Sec:summary}, we summarize our findings and discuss 
potential future directions. 

\section{Generalized uncertainty principle}
\label{Sec:model}

The most widely studied generalized uncertainty principle leading to a minimal length is of the form \cite{Das:2008kaa}: 
\begin{equation}
\Delta Q \Delta P \geq \frac{\hslash}{2}  (1+\beta \Delta P^2), \label{GUP}
\end{equation}
where $\Delta Q$ and $\Delta P$ are uncertainties on the generalized 
position and momenta, respectively. Here, $\beta$  is a small dimensional 
parameter, serving as a measure of the strength of quantum gravity effects. 
This parameter is assumed to be positive definite here. However, negative 
values of $\beta$ have also been studied in the literature \cite{Bojowald:2011jd,Ong:2018zqn}.

Importantly, in this article, the framework of field theory is considered, 
for which $[Q]=E^{-1/2}$, $[P]=E^{1/2}$ and as a consequence $[\beta]= E^{-1}$.
Since the modification is considered to be due to Planck-scale physics, 
it is therefore expected that $\beta \sim 1/E_{\text{Pl}}$, where the Planck 
energy $E_{\text{Pl}} \approx 1.22 \cdot 10^{19}$ GeV. Notably, this field theory scenario differs substantially from the usual case of a point particle, for which  $[Q]=E^{-1}$, $[P]=E$, and consequently 
$[\beta]= E^{-2}$, leading to the $\beta$ parameter being $\beta \sim 1/E_{\text{Pl}}^2$. 
The differing dimensions of the canonical variables between both scenarios, 
therefore, suggest that the effects of new physics at the Planck scale transcribed 
by the GUP could have a more pronounced impact in the field theory context 
(where they are expected to be suppressed by $E/E_\text{Pl}$ to some power) 
compared to the point particle scenario (where they are expected to be suppressed by $E^2/E_\text{Pl}^2$ to some power).

Additionally, while the GUP in the 
context of a point particle implies the existence of a minimal physical 
length, the situation differs in the field theoretical framework considered 
here. In this setting, the GUP does not give rise to a minimal length, 
but rather suggests a minimal field value.

The GUP given by Eq. (\ref{GUP}) can be derived from the deformed 
commutation relation:
\begin{equation}
[ \hat{Q}, \hat{P}]= i \hslash (\hat{\mathbb{I}} +\beta \hat{P}^2). \label{Com1}
\end{equation}
This commutation relation is predicted by different 
Planck-scale physics models, such as relative locality 
\cite{Amelino-Camelia:2011lvm} or loop quantum gravity \cite{Bojowald:2011jd,Mielczarek:2013rva}. 

An observation made in \cite{Pedram:2011aa} is that this commutation relation (\ref{Com1}) transforms 
into the standard one $[\hat{q},\hat{p}] = i \hslash \mathds{1}$ under the following change of 
variables: 
\begin{align}
\hat{Q} &= \hat{q}, \label{defQ} \\
\hat{P} &=  \frac{\tan (\sqrt{\beta} {\hat{p}})}{\sqrt{\beta}}. \label{defP}
\end{align}
This can be proven using the fact that for any function $f(\hat{p})$,  
$[\hat{q},f(\hat{p})] = i \hslash \frac{df(\hat{p})}{d\hat{p}}$ if $[\hat{q},\hat{p}] 
= i \hslash \hat{\mathbb{I}}$. Since the commutation relation changes, the above change 
of variables is not a canonical transformation. 
   
This article aims to consider the quantum properties of light, taking 
into account the deformed commutation relation (\ref{Com1}). Some   
studies in this direction have already been made in Refs. \cite{Bosso:2017ndq,Conti:2018hhj}.
 
The simplest case we are going to begin with is a single-mode light for which 
the standard Hamiltonian takes the form
\begin{equation}
\hat{H} = \frac{1}{2} \left( \hat{P}^2 +\omega^2 \hat{Q}^2  \right),
\label{SingleModeH}
\end{equation}
where $\omega$ denotes the frequency of the mode, 
so that $[\omega]= E$. Consequently, $[\beta \omega \hbar ] = 1$.
Importantly, the polarization states of light are not 
considered here but the amplitude of the field solely. 

Employing the change of variables (\ref{defQ}) and (\ref{defP}) and 
expanding the obtained expression up to the linear order in $\beta$ leads to 
\begin{equation}
\hat{H} = \frac{1}{2} \left( \hat{p}^2 +\omega^2 \hat{q}^2  \right) + \frac{\beta}{3} \hat{p}^4 + \mathcal{O}(\beta^2).
\label{FullHam}
\end{equation}
Worth mentioning is that only even powers of $\hat{p}$
are contributing to the series. From now on all $\mathcal{O}(\beta^2)$ contributions will be neglected
and we will focus only on the leading effect . 

For further convenience, the full Hamiltonian 
(\ref{FullHam}) is decomposed into a free 
$\hat{H}_0 =  \frac{1}{2} \left( \hat{p}^2 +\omega^2 \hat{q}^2  \right)
$, and an interaction
   $\hat{H}_1 =  \frac{\beta}{3} \hat{p}^4$
parts, such that
\begin{equation}
\hat{H} = \hat{H}_0+ \hat{H}_1. \label{HamDecomp}
\end{equation}

At this point, it is useful to introduce the standard creation and annihilation operators 
$\hat{a}^{\dagger}$ and $\hat{a}$ defined in the usual way:
\begin{align}
\hat{q} &:= \sqrt{\frac{\hslash}{2\omega}} \left(\hat{a}^{\dagger} +\hat{a}\right), \\ 
\hat{p} &:= i \sqrt{\frac{\hslash \omega}{2}} \left(\hat{a}^{\dagger} -\hat{a}\right),
\end{align} 
so that $[\hat{a},\hat{a}^{\dagger}]= \hat{\mathbb{I}}$. The free Hamiltonian 
then reads $\hat{H}_0 = \frac{\hslash \omega}{2} \left( \hat{a}^\dagger \hat{a} 
+ \hat{a} \hat{a}^\dagger \right) $ and verifies $[\hat{H}_0,\hat{a}^\dagger]=\hslash \omega \hat{a}^\dagger$ and $[\hat{H}_0,\hat{a}]=-\hslash \omega \hat{a}$. 
Furthermore, the interaction Hamiltonian reads
\begin{equation}
\hat{H}_1 = \frac{\beta}{3}\left(\frac{\hslash \omega}{2} \right)^2 
\left( \hat{a}^{\dagger}-\hat{a}\right)^4.   
\end{equation}

By applying the standard perturbation theory, one can find that at the  
first order of the perturbative expansion, the Hamiltonian (\ref{HamDecomp})
eigenvalues are
\begin{align}
    E_n^{(1)} &= \bra{n^{(0)}}\hat{H}_0 + \hat{H}_1 \ket{n^{(0)}} \nonumber \\
    & = \bra{n} \hslash \omega \left( \hat{N} + \frac{1}{2} \hat{\mathbb{I}} \right) \ket{n}+
    \bra{n}\frac{\beta}{3} \hat{p}^4 \ket{n} \nonumber \\
    &= \hslash \omega \left(n+ \frac{1}{2}\right) + \beta \frac{\hslash^2 \omega^2}{4} \left(2n^2 + 2n +1 \right),
\end{align}
in which $\hat{N} := \hat{a}^\dag \hat{a}$. One easily recovers 
from this expression that the difference of energy levels at zeroth 
order in $\beta$ is $E_n^{(0)}-E_m^{(0)}=\hslash \omega (n-m)$.

The associated eigenstates are, at first order in $\beta$,
\begin{align}
\ket{n^{(1)}} &:= \ket{n}+\sum_{m\neq n} 
\frac{\bra{m}\hat{H}_1\ket{n}}{E_n^{(0)}-E_m^{(0)}} \ket{m}  \nonumber \\ 
&=\ket{n}+ \frac{\beta \hslash \omega}{12}\left[ 
- \frac{1}{4}\sqrt{\frac{(n+4)!}{n!}} \ket{n+4} \right. \nonumber\\
&+\sqrt{n+1}\sqrt{n+2}\left(2n+3\right)\ket{n+2}  \nonumber\\
&- \sqrt{n}\sqrt{n-1} \left(2n-1\right)\ket{n-2}  \nonumber\\
&+\left. \frac{1}{4}\sqrt{\frac{n!}{(n-4)!}}\ket{n-4} \right] .
\end{align}

In particular, the first-order vacuum energy is 
\begin{align}
    E_0^{(1)} = \frac{\hslash \omega}{2}\left(1+\beta \frac{\hslash \omega}{2}\right),
\end{align}
and the first-order vacuum state
\begin{equation}
\ket{0^{(1)}}= \ket{0}+\frac{\beta \hslash \omega}{12}\left( 
3\sqrt{2}\ket{2}  - \sqrt{\frac{3}{2}} \ket{4} 
\right).
\label{01state}
\end{equation}
Therefore, the GUP correction slightly leverages the 
ground state energy. 

\section{Time evolution} 
\label{Sec:evolution}

The time evolution of a single-mode light can
be studied by introducing the following operator:
\begin{equation}
\hat{F}(t) : = \hat{U}^{-1}_0(t) \hat{U}(t),
\end{equation}
where the unitary operator $ \hat{U}_0(t) := \exp\left(- \frac{i}{\hslash} \hat{H}_0 t\right)$ 
utilizes the free part of the Hamiltonian (\ref{HamDecomp}), whereas 
$ \hat{U}(t) = \exp\left(- \frac{i}{\hslash}  \hat{H} t\right)$.  
The operator $\hat{F}$ satisfies the equation
\begin{equation}  
\frac{d\hat{F}(t)}{dt} =  - \frac{i}{\hslash}  \hat{H}_1^I(t)\hat{F}(t), 
\end{equation}
which has a solution in the form of a Dyson series:
\begin{align}
\hat{F}(t) &=  \hat{T} \exp \left(- \frac{i}{\hslash} \int_0^t \hat{H}_1^I(t')dt'\right) \nonumber \\
               &=\hat{\mathbb{I}}- \frac{i}{\hslash} \int_0^t \hat{H}_1^I(t')dt'+\mathcal{O}(1/\hslash^2), 
\end{align}
where 
\begin{equation}
\hat{H}_1^I(t) :=  \hat{U}^{-1}_0(t) \hat{H}_1  \hat{U}_0(t) 
\end{equation}  
is the interaction Hamiltonian in the interaction picture. 
Furthermore, $\hat{T}$ is the time ordering operator, and the 
operator $\hat{F}(t)$ satisfies the initial condition $\hat{F}(0)
=\hat{\mathbb{I}}$.

The time evolution of an initial state $|\Psi(0) \rangle$ is given by 
\begin{align}
|\Psi(0) \rangle =&\ \hat{U}(t) |\Psi(0) \rangle \nonumber \\
=&\ \hat{U}_0(t)\hat{F}(t)|\Psi(0) \rangle \nonumber \\
=&\ \hat{U}_0(t)|\Psi(0) \rangle \nonumber \\
&- \frac{i}{\hslash} \hat{U}_0(t)\int_0^t \hat{H}_1^I(t')dt'|\Psi(0) \rangle 
+ \mathcal{O}(1/\hslash^2). \label{StateEvol}
\end{align}

From now on, all $\mathcal{O}(1/\hslash^2)$ contributions will be neglected and we will focus on the leading effect only. 

Employing the Baker-Campbell-Hausdorff formula, one can find that
\begin{align}
\hat{U}^{-1}_0(t) \hat{a}^{\dagger}  \hat{U}_0(t) &= \hat{a}^{\dagger} e^{i\omega t},          \\
\hat{U}^{-1}_0(t) \hat{a} \hat{U}_0(t) &=  \hat{a} e^{-i\omega t}.  
\end{align}

The use of the above leads to

\begin{equation}
\hat{H}_1^I(t) = \frac{\beta}{3}\left(\frac{\hslash \omega}{2} \right)^2 
\left( \hat{a}^{\dagger}e^{i\omega t} -\hat{a}e^{-i\omega t} \right)^4,   
\end{equation}

for $\hat{H}_1$ given by the interaction term in Eq. \eqref{FullHam}. \\

When developing this expression and using the commutation 
relation between the creation and annihilation operators, 
one gets the following form of the interaction Hamiltonian 
in the interaction picture\footnote{It may be useful to observe that 
\begin{align}
    \hat{a}\left(\hat{a}^\dagger\right)^n &= j \left(\hat{a}^\dagger\right)^{n-1} + \left(\hat{a}^\dagger\right)^j \hat{a} \left(\hat{a}^\dagger\right)^{n-j} \,,\\
    \left(\hat{a}\right)^n \hat{a}^\dagger &= j \left(\hat{a}\right)^{n-1} + \left(\hat{a}\right)^{n-j} \hat{a}^\dagger \left(\hat{a}\right)^j
    \,,
\end{align}
for all integers $j\in[0,n]$.}:
\begin{align}
    \hat{H}_1^I (t) =& \frac{\beta \hslash^2 \omega^2}{12} \big[ \left(\hat{a}^\dagger\right)^4 e^{4i\omega t} \label{H1It} \nonumber \\
    &- \left(4 \left(\hat{a}^\dagger\right)^3\hat{a} + 6 \left(\hat{a}^\dagger\right)^2 \right) e^{2i\omega t} \nonumber \\
    &+ \left(6 \left(\hat{a}^\dagger\right)^2 \left(\hat{a}\right)^2 + 12 \hat{a}^\dagger \hat{a} + 3 \hat{\mathbb{I}} \right) \nonumber\\
    &- \left(4 \hat{a}^\dagger \left(\hat{a}\right)^3 + 6 \left(\hat{a}\right)^2\right) e^{-2i\omega t} \nonumber \\
    &+ \left(\hat{a}\right)^4 e^{-4i\omega t} \big]  \,. 
\end{align}

The time integral contributing to the series formula (\ref{StateEvol}) can be analytically evaluated and decomposed into real 
and imaginary parts:
\begin{equation}
    \int_0^t \hat{H}_1(t') dt' = \beta \frac{\hslash^2 \omega}{24} \left( \hat{\mathcal{R}}(t) + i \hat{\mathcal{I}}(t) \right),
\end{equation}
with
\begin{align}
    \hat{\mathcal{R}}(t) =&  2 \omega t \left[ 6 \left( \hat{a}^\dag \right)^2 \left( \hat{a} \right)^2 + 12 \hat{a}^\dag \hat{a} + 3 \hat{\mathbb{I}} \right] \\ \nonumber
    +& \frac{1}{2} \sin \left( 4 \omega t \right) \left[ \left( \hat{a}^\dag \right)^4 + \left( \hat{a}^4 \right) \right] \\ \nonumber
    - & \sin \left( 2 \omega t \right) \left[ 4 \left( \hat{a}^\dag \right)^3 \hat{a} + 6 \left( \hat{a}^\dag \right)^2  + 4 \hat{a}^\dag \left( \hat{a} \right)^3 + 6 \left( \hat{a} \right) ^2 \right],  \nonumber
\end{align}
and 
\begin{align}
    \hat{\mathcal{I}}(t) =& \frac{1}{2} \left( 1 - \cos \left( 4 \omega t \right) \right) \left[ \left( \hat{a}^\dag \right)^4 - \left( \hat{a}^4 \right) \right] \\ \nonumber
    +& \left( 1 - \cos \left( 2 \omega t \right) \right) \\ \nonumber
    \times& \left[ - 4 \left( \hat{a}^\dag \right)^3 \hat{a} - 6 \left( \hat{a}^\dag \right)^2  + 4 \hat{a}^\dag \left( \hat{a} \right)^3 + 6 \left( \hat{a} \right) ^2 \right]. \nonumber
\end{align}

\section{Squeezing} 
\label{Sec:squeezing}

In the phase-space formulation of quantum mechanics the 
system is described by a quantum state in a space defined 
by generalized position and momentum coordinates, known 
as phase space. Quantum squeezing corresponds in this context to the 
distortion of the quantum state, represented, e.g., by its Wigner 
quasiprobability function $W(q,p)$, in this phase space. 
More specifically, it refers to the reduction/spread of the uncertainty in one direction (say position) at the expense 
of increasing/decreasing it in the conjugate dimension (momentum), 
ensuring compliance with the Heisenberg or more generally the
Robertson-Schr\"odinger uncertainty relation. The squeezed 
state is represented as an ellipse rather than a circle 
(the latter corresponding to equal uncertainty in both dimensions) 
in the phase space. Worth emphasizing is that quantum 
squeezing is of particular interest in quantum optics 
and quantum information science, where it can improve 
measurement precisions and information processing 
capabilities by reducing quantum noise and enhancing signal 
strength in a specific direction (see, e.g., Refs. \cite{Lawrie,Frascella}).

The squeezing prompted by the GUP
given by Eq. \eqref{GUP} is examined in the next section, focusing 
on two classes of states: quantum harmonic oscillator energy eigenstates 
$\ket{n}$ and Glauber's coherent states $\ket{\alpha}$, which play 
key roles in quantum optics and quantum information sciences. 

Squeezing properties are quantified by
the first and second moments of the $\hat{q}$ 
and $\hat{p}$ operators. We introduce
\begin{align}
\langle \hat{q} \rangle &:= \bra{\Psi(t)} \hat{q} \ket{\Psi(t)}, \\
\langle \hat{p} \rangle &:= \bra{\Psi(t)} \hat{p} \ket{\Psi(t)}, \\
\Delta \hat{q} &:= \sqrt{\bra{\Psi(t)} \hat{q}^2 \ket{\Psi(t)} - \left( \bra{\Psi(t)} \hat{q}  \ket{\Psi(t)} \right)^2}, \\
\Delta \hat{p} &:= \sqrt{\bra{\Psi(t)} \hat{p}^2 \ket{\Psi(t)} - \left( \bra{\Psi(t)} \hat{p}  \ket{\Psi(t)} \right)^2}, \\
C_{qp} &:= \bra{\Psi(t)}(\hat{q}-\langle \hat{q} \rangle)(\hat{p}-\langle \hat{p} \rangle)  \ket{\Psi(t)}_{\text{Weyl}} \nonumber \\
    &=\frac{1}{2} \bra{\Psi(t)}(\hat{q}\hat{p}+\hat{p}\hat{q})\ket{\Psi(t)}- \langle \hat{q} \rangle \langle \hat{p} \rangle,
\end{align}
so that the Robertson-Schr\"odinger uncertainty principle holds:
\begin{equation}
(\Delta \hat{q})^2(\Delta \hat{p})^2-C_{qp}^2 \geq \hslash^2/4. 
\label{RSU}
\end{equation}
In the definition of the covariance $C_{qp}$ the Weyl
symmetrization is applied. Furthermore, we will add superscript 
$^{(0)}$ in case of the zeroth-order formulas ($\beta \rightarrow 0$). 

Alternatively, the  Robertson-Schr\"odinger uncertainty can be written as 
\begin{equation}
(\Delta \hat{q})^2(\Delta \hat{p})^2(1-\rho^2) \geq \hslash^2/4, 
\label{RSUrho}
\end{equation}
where $\rho$ is the  dimensionless correlation coefficient
\begin{equation}
\rho :=  \frac{C_{qp}}{\Delta \hat{q} \Delta \hat{p}}.
\end{equation}
When $\rho$ is different from zero, the semiaxes of the ellipsoid 
of covariance do not overlap with the directions $q$ and $p$.

To quantify this effect, it is convenient to introduce the 
covariance matrix
\begin{equation}
{\bf \Sigma} :=  \left[\begin{array}{cc} \omega ( \Delta  \hat{q})^2 & C_{qp} \\ C_{qp}& ( \Delta \hat{p})^2/\omega\end{array}  \right],
\label{CovarianceMatrix}
\end{equation}
where the $\omega$ factor has been introduced for dimensional reasons. 
The eigenvalues of the  previous matrix are
\begin{eqnarray}
\lambda_{\pm} = \frac{1}{2}\left[ \text{tr}  {\bf \Sigma} \pm \sqrt{(\text{tr}{\bf \Sigma})^2 -4 \det {\bf \Sigma} } \right],
\label{eigenvaluesLambda}
\end{eqnarray}
where $\text{tr} {\bf \Sigma}= \omega(\Delta \hat{q})^2 +( \Delta
\hat{p})^2/\omega$ and $\det {\bf \Sigma}=( \Delta \hat{q})^2(\Delta
\hat{p})^2(1-\rho^2)$. Because of the square root in Eq. (\ref{eigenvaluesLambda}), 
the $\mathcal{O}(\beta^2)$ factors could in principle bring a contribution
of the $\beta$ order in the squeezing amplitude. However, the terms of the 
order $\beta^2$ in $(\Delta \hat{q})^2$, in $(\Delta \hat{p})^2$, and in $C_{qp}$ 
bring no contribution of the order $\beta^2$ in $\left[(\text{tr}{\bf \Sigma})^2
-4 \det {\bf \Sigma}\right]$ whatever the state considered (see Appendix A). 
In consequence, the unknown factors do not contribute to the linear in 
$\beta$ expressions for the eigenvalues $\lambda_{\pm}$.

Importantly, the square roots of the eigenvalues have 
interpretations of major and minor diameters of the ellipsoid 
of covariance, respectively, so that the uncertainty 
relation (\ref{RSU}) takes the form
\begin{equation}
\sqrt{\lambda_{+}}\sqrt{\lambda_{-}} 
\geq \hslash/2.
\end{equation}

In the eigenframe the correlation vanishes, and the relative 
values of $\sqrt{\lambda_{+}}$ and $\sqrt{\lambda_{-}}$ can be 
used to quantify the squeezing of the state. Specifically, after
suitable normalization of the variables, we can write
\begin{eqnarray}
\sqrt{\lambda_{+}} &=& \sqrt{\frac{\hslash}{2}} e^r,\\
\sqrt{\lambda_{-}} &=& \sqrt{\frac{\hslash}{2}} e^{-r}, 
\label{lambdar}
\end{eqnarray}
where  $r$ is the squeezing amplitude related to the 
complex squeezing parameter $\xi = |r| e^{i \gamma}$.
Geometrically, $\gamma/2$ is the angle between the minor
axes of the ellipsoid and the $q$ axis. The $\xi$ 
parameter enters the squeezing operator $\hat{S}(\xi)$
as follows:
\begin{equation}
\hat{S}(\xi) := \exp 
\left( \frac{1}{2}( \xi^* \hat{a}^2
-\xi  \hat{a}^{\dagger 2} )  \right).
\end{equation}

\subsection{Squeezing of quantum harmonic oscillator eigenstates:  $|\Psi(0) \rangle = | n \rangle $}

The free evolution of a $\ket{n}$ state under the harmonic oscillator Hamiltonian is given by

\begin{equation}
    \hat{U}_0 |n\rangle = e^{- i\omega t \left(n+\frac{1}{2}\right)} |n\rangle\,.
\end{equation}

To compute the evolution under the full Hamiltonian at 
first order in $\beta$, one needs to use
\begin{align}
\hat{H}_1^I|n \rangle =&\frac{\beta \hslash^2 \omega^2}{12} \big[ e^{4i\omega t} \sqrt{n+1} \sqrt{n+2} \sqrt{n+3} \sqrt{n+4} \, |n+4 \rangle \nonumber \\
&-e^{2i\omega t} \sqrt{n+1} \sqrt{n+2} \left(4n +6\right) \, |n+2 \rangle \nonumber \\
&+ \left( 6n^2 + 6n   +3\right) \,|n \rangle \nonumber \\
&- e^{-2i\omega t} \sqrt{n} \sqrt{n-1} \left(4n-2\right)  \, |n-2 \rangle \nonumber \\
&+ e^{-4i\omega t} \sqrt{n} \sqrt{n-1} \sqrt{n-2} \sqrt{n-3} \,|n-4 \rangle \big]\,,
\end{align}
obtained from Eq. \eqref{H1It}. Recall that $n \geq 0 $ for the 
initial state to be well defined. Similarly, any state 
$\ket{n-m}$ is well defined if and only if $(n-m) \geq 0$.

Under the evolution given Eq. (\ref{StateEvol}) the state 
$\ket{n}$ at any time boils down to
\begin{widetext}
\begin{align}
    |\Psi(t)\rangle &=  e^{- i\omega t \left(n+\frac{1}{2}\right)} |n\rangle\ - \frac{\beta \hslash \omega}{24} \left[ \frac{1}{2} \left(e^{-i\omega t \left( n+\frac{1}{2} \right)} - e^{-\frac{i\omega t}{2}\left(2n+9\right)}\right) \frac{\sqrt{\left(n+4\right)!}}{\sqrt{n!}} \, |n+4 \rangle \right.  \label{PsiTimeN}\\
&-\left(e^{-i\omega t \left( n+\frac{1}{2} \right)} - e^{-\frac{i\omega t}{2}\left(2n+5\right)}\right) \frac{\sqrt{\left(n+2\right)!}}{\sqrt{n!}} \left(4n+6\right) \, |n+2 \rangle 
+ 2 i \omega t e^{- i\omega t \left( n+\frac{1}{2}\right)} \left(6n^2 +6n+3\right) \,|n \rangle \nonumber \\
&+ \left(e^{-i\omega t \left( n+\frac{1}{2} \right) } - e^{-\frac{i\omega t}{2}\left(2n-3\right)}\right) \frac{\sqrt{n!}}{\sqrt{\left(n-2\right)!}} \left(4n-2\right)\, |n-2 \rangle \left. - \frac{1}{2} \left( e^{-i\omega t \left( n+\frac{1}{2} \right)} - e^{-\frac{i\omega t}{2}\left(2n-7\right)} \right) \frac{\sqrt{n!}}{\sqrt{\left(n-4\right)!}} \,|n-4 \rangle \right]\,. \nonumber
\end{align}
\end{widetext}

 It should be emphasized here that the normalization $\braket{\Psi|\Psi} = 1 + \mathcal{O} 
(\beta^2)$ differs from unity at second order in $\beta$ only, not 
at the $\mathcal{O}(\beta)$ order. 

For an evolution governed by the corrected Hamiltonian $\hat{H}$,
the $q$ and $p$ first moments, computed using the state (\ref{PsiTimeN}), vanish at any time:

\begin{equation}
    \langle \hat{q} \rangle =  0 = \langle \hat{p} \rangle.
\end{equation}
Because of to the $\hat{p}^4$-type of the interaction term, this is 
satisfied at any order in $\beta$. For the same reason, 
any odd power of $\hat{q}$ and $\hat{p}$, including the mixed 
terms, will also vanish.  

The mean value of the $\ket{n}$ state in the phase space is 
therefore not modified by the new dynamics. Its dispersions 
in $q$ and $p$, however, vary as
\begin{align}
    &(\Delta \hat{q})^2 = \frac{\hslash}{2\omega} \left(1+2n\right) \nonumber\\
    &+ \beta \hslash^2 \sin^2\left(\omega t\right) 
    \left(2n^2 + 2n +1\right) +\mathcal{O} \left( \beta^2 \right)\,, \\
    &(\Delta \hat{p})^2  = \frac{\hslash\omega}{2} \left(1+2n\right) \nonumber\\
    &- \beta \hslash^2\omega^2 \sin^2\left(\omega t\right) 
\left(2n^2 + 2n +1\right) +\mathcal{O} \left( \beta^2 \right) \,.
\end{align}

The covariance evaluated at any time writes 
\begin{align}
    C_{qp} = \frac{\beta \hslash^2\omega}{2} \sin\left(2\omega t\right) \left(2n^2 + 2n +1\right) 
     +\mathcal{O} \left( \beta^2 \right).
\end{align}

One can easily verify that the leading order in the $\beta$ 
contribution to the Robertson-Schr\"odinger relation 
cancels out so that 
\begin{equation}
\frac{\hslash^2}{4}(1+2n)^2+ \mathcal{O} \left( \beta^2 \right)\geq \frac{\hslash^2}{4}. 
\end{equation}

Importantly, the first order in $\beta$ corrections to 
the quadratic moments of $\hat{q}$ and $\hat{p}$
oscillate in time at the frequency $\omega$.
There is, therefore, no accumulative contribution to 
the dispersions of the state at this order.

For the case under consideration, the square roots 
of the eigenvalues of the covariance matrix (\ref{CovarianceMatrix}) can be expressed as follows:
\begin{align}
\sqrt{\lambda_{+}} &= \sqrt{\frac{\hslash}{2}} \sqrt{1+2n}e^r, \\
\sqrt{\lambda_{-}} &= \sqrt{\frac{\hslash}{2}}\sqrt{1+2n} e^{-r}, 
\end{align}
where the squeezing amplitude $r$ for an arbitrary state $\ket{n}$ is given by
\begin{equation}
    r = \frac{1+2n+2n^2}{1+2n} \beta \hslash \omega |\sin\left(\omega t\right)| + \mathcal{O}\left(\beta^2\right)\,. 
    \label{Squeezing-nState}
\end{equation}

Consequently, for large $n$, the squeezing amplitude
grows as $\mathcal{O}\left(n\right)$.

For the vacuum state $\ket{\Psi(0)}=\ket{0}$ the
dispersions in $q$ and $p$ and the covariance reduce to
\begin{align}
(\Delta \hat{q})^2 & = \frac{\hslash}{2\omega}\left[1+2 \beta \hslash \omega  \sin^2\left(\omega t\right) +\mathcal{O} \left( \beta^2 \right)\right], \label{Dqvac} \\
(\Delta \hat{p})^2 & = \frac{\hslash\omega}{2}\left[1- 2\beta \hslash\omega \sin^2\left(\omega t\right) +\mathcal{O} \left( \beta^2 \right)\right], \label{Dpvac}\\
C_{qp} & = \beta \frac{\hslash^2\omega}{2} \sin\left(2\omega t\right) +\mathcal{O} \left( \beta^2 \right), \label{Cqpvac}
\end{align}
and the squeezing amplitude Eq. \eqref{Squeezing-nState} boils down to
\begin{equation}
    r = \beta \hslash \omega |\sin\left(\omega t\right)| + \mathcal{O} \left( \beta^2\right).
\end{equation}

The maximal squeezing magnitude 
expected for the corrected vacuum state at first order in 
$\beta$ expansion is therefore
\begin{equation}
r_{\text{max}, \ket{0}} \approx  \beta \hslash \omega.
\label{rmaxvacuum}
\end{equation}

\subsection{Squeezing of Glauber's coherent states: $|\Psi(0) \rangle = | \alpha \rangle $}

The second example we are going to consider is a state that is initially Glauber's coherent state of a harmonic oscillator:

\begin{equation}
|\Psi(0) \rangle=|\alpha \rangle := e^{- \frac{1}{2} |\alpha|^2}\sum_{n=0}^{\infty} \frac{\alpha^n}{\sqrt{n!}}| n \rangle, 
\end{equation}
introduced such that $\hat{a}|\alpha \rangle  = \alpha |\alpha \rangle $, 
where $\alpha = |\alpha|e^{i\theta} \in \mathbb{C}$. 
Let us remind the reader that the free evolution of such a state is given by
\begin{equation}
|\Psi^{(0)}(t) \rangle :=\hat{U}_0(t)|\alpha \rangle = e^{-i \frac{\omega t}{2}} | \alpha e^{-i\omega t} \rangle.  
\end{equation}

In this state, the $\hat{q}$ and $\hat{p}$ mean values are
\begin{align}
    \langle \hat{q} \rangle^{(0)} & = |\alpha|  \sqrt{\frac{2 \hbar }{\omega }} \cos (\theta -\omega t ),\\
   \langle \hat{p} \rangle^{(0)} & = - |\alpha|  \sqrt{2 \omega  \hbar } \sin (\theta -\omega t ), 
\end{align}
and the standard deviations and the covariance write
\begin{align}
\Delta \hat{q}^{(0)} &= \sqrt{\frac{\hbar}{2 \omega}}, \\
\Delta \hat{p}^{(0)} &= \sqrt{\frac{\hbar \omega}{2}}, \\
C_{qp}^{(0)} &= 0, 
\label{dispersions free evolution}
\end{align}
which are constant in time.

The right-hand action of the interaction Hamiltonian on the coherent state
$\ket{\alpha}$ is given by
\begin{align}
    \hat{H}_1^I (t) \ket{\alpha} =& \frac{\beta \hslash^2 \omega^2}{12} \big[ \left(\hat{a}^\dagger\right)^4 e^{4i\omega t} \nonumber \\
    &- \left(4 \alpha \left(\hat{a}^\dagger\right)^3 + 6 \left(\hat{a}^\dagger\right)^2 \right) e^{2i\omega t} \nonumber \\
    &+ \left(6 \alpha^2\left(\hat{a}^\dagger\right)^2  + 12 \alpha\hat{a}^\dagger  + 3 \,\mathds{1} \right) \nonumber\\
    &- \left(4 \alpha^3 \hat{a}^\dagger  + 6 \alpha^2 \right) e^{-2i\omega t} \nonumber \\
    &+ \left(\alpha\right)^4 e^{-4i\omega t} \big]\ket{\alpha}  , 
    \label{HIntAlpha}
\end{align}
which allows one to evaluate the mean values of $\hat{q}$ and $\hat{p}$ operators:
\begin{widetext}
\begin{align}
   \langle \hat{q} \rangle & = |\alpha|  \sqrt{\frac{2\hbar }{\omega }} \cos (\theta -\omega t )+\frac{1}{6 \sqrt{2}}|\alpha|  \beta  \omega  \hbar  \sqrt{\frac{\hbar }{\omega }} \big(12 (|\alpha| ^2+1) \omega t  \sin (\theta ) \cos (\omega t )-2 \sin (\omega t ) \big[-6 \sin (\theta )    \nonumber  \\
   & +|\alpha| ^2 (-6 \sin (\theta )+\sin (3 \theta )+3 \sin (3 \theta -2 \omega t ))+6 (|\alpha| ^2+1) \omega t  \cos (\theta )\big]\big)+\mathcal{O} \left( \beta^2 \right), \label{averalphaq} \\
   \langle \hat{p} \rangle & = -|\alpha|  \sqrt{2 \omega  \hbar } \sin (\theta -\omega t )+\frac{1}{3} \sqrt{2} |\alpha|   \beta  (\omega  \hbar )^{3/2} \big[|\alpha|  ^2 \sin ^2(\omega t ) \sin (3 \theta -\omega t ) \nonumber \\
   &-3 (|\alpha|  ^2+1) (\omega t  \cos (\theta -\omega t )-\cos (\theta ) \sin (\omega t ))\big]+\mathcal{O} \left( \beta^2 \right).\label{averalphap}
\end{align}
\end{widetext}

A displacement of the 
mean values of $\hat{q}$ and $\hat{p}$ due to the 
GUP correction is exhibited. Importantly, the correction is not 
only oscillatory in time but, due to the multiplicative 
$\omega t$ factor, a time-cumulative effect appears.

The phase factor $\theta$ plays no important role in our discussion and thus can be fixed at $\theta=0$ for the simplicity 
of the further analysis. Then, at late times the formulas 
(\ref{averalphaq}) and (\ref{averalphap}) are well approximated by
\begin{widetext}
\begin{align}
   \langle \hat{q} \rangle & \approx |\alpha|  \sqrt{\frac{2\hbar }{\omega }} \left[ \cos (\omega t )
   -  \beta  \omega  \hbar  (\omega t )    (|\alpha| ^2+1)\sin (\omega t )  \right] ,  \\
   \langle \hat{p} \rangle & \approx  |\alpha|  \sqrt{2 \omega  \hbar } \left[ \sin (\omega t )-   \beta \omega  \hbar   (\omega t ) (|\alpha|  ^2+1)  \cos (\omega t ) \right].
\end{align}
\end{widetext}
The expectation value of the annihilation operator $\hat{a}$ 
(which gives the mean location of the state on the complex plane representation of the 
phase space) becomes
\begin{align}
\langle \hat{a} \rangle &= \sqrt{\frac{2\omega}{\hslash}}\langle \hat{q} \rangle+i\sqrt{\frac{1}{2\hslash \omega}} \langle \hat{p} \rangle  \nonumber \\
& \approx |\alpha|(1-\beta \omega  \hbar   (\omega t ) (|\alpha|  ^2+1) \sin (2\omega t))e^{i \varphi}, 
\end{align}
where
\begin{equation}
\varphi \approx \omega t \left[1-2\beta \omega  \hbar   (\omega t)(|\alpha|  ^2+1)  \right].
\end{equation}

 At leading order in $\beta$, the displacement of the state from the origin of the phase space follows oscillations whose amplitude grows linearly in time.\\ 

The variances and the covariance of the canonically conjugated variables $q$ and $p$ can 
be evaluated with the use of Eq. \eqref{HIntAlpha}, leading to
\begin{widetext}
\begin{align}
\left(\Delta \hat{q}\right)^2 &= \frac{\hbar }{2 \omega }-\frac{1}{6} \beta  (\hbar ^2 (|\alpha| ^2 (|\alpha|  (-9 \cos (\theta )-4 |\alpha|  \sin (\omega t ) (\sin (4 \theta -3 \omega t )-|\alpha|  \sin (5 \theta -3 \omega t ))+6 |\alpha|  \omega t  \sin (2 \theta -2 \omega t ) \nonumber \\
 &+9 \cos (\theta -2 \omega t ))-3 \cos (2 \omega t ) (2 \omega t  \sin (2 \theta )+1)+3 \sin (2 \omega t ) (\sin (2 \theta )+2 \omega t  \cos (2 \theta )+\omega t )) \nonumber\\
 &+3 (|\alpha| ^2+\cos (2 \omega t )-1)))+\mathcal{O} \left( \beta^2 \right), \label{CoherentSecondMoments1} \\
\left(\Delta \hat{p}\right)^2 &= \frac{\hbar \omega }{2 }+\frac{1}{6} \beta \omega^2  (\hbar ^2 (|\alpha| ^2 (|\alpha| (-9 \cos (\theta )-4 \alpha  \sin (\omega t ) (\sin (4 \theta -3 \omega t )-|\alpha| \sin (5 \theta -3 \omega t ))+6 |\alpha|  \omega t  \sin (2 \theta -2 \omega t ) \nonumber \\
 &+9 \cos (\theta -2 \omega t ))-3 \cos (2 \omega t ) (2 \omega t  \sin (2 \theta )+1)+3 \sin (2 \omega t ) (\sin (2 \theta )+2 \omega t  \cos (2 \theta )+\omega t )) \nonumber \\
 &+3 (|\alpha| ^2+\cos (2 \omega t )-1)))+\mathcal{O} \left( \beta^2 \right),\label{CoherentSecondMoments2} \\
C_{qp} &= -\frac{1}{6} \beta  \omega  \hbar ^2 (-4 |\alpha| ^4 \sin (\omega t ) \cos (4 \theta -3 \omega t )+6 (2 |\alpha| ^2+3) |\alpha| ^2 \omega t  \cos (2 \theta -2 \omega t ) \nonumber \\
& +(|\alpha|^2 \left(|\alpha| ^2+4\right) \cos (2 \theta )-3 \left(2 |\alpha| ^4+4 |\alpha| ^2+1\right)) \sin (2 \omega t ))+\mathcal{O} \left( \beta^2 \right). 
\label{CoherentSecondMoments3}
\end{align}
\end{widetext}

In the $|\alpha|\rightarrow 0$ limit the case of the 
vacuum state $\ket{0}$, for which Eqs. \eqref{Dqvac}, \eqref{Dpvac}, and \eqref{Cqpvac} hold, 
is correctly recovered. Furthermore, the Robertson-Schr\"odinger relation gains 
no correction in the order linear in $\beta$, i.e.,
\begin{equation}
\frac{\hslash^2}{4}+\mathcal{O} \left( \beta^2 \right)\geq \frac{\hslash^2}{4}. 
\end{equation}

By applying the formula (\ref{lambdar}) and the results of 
Appendix A, the squeezing amplitude $r$ for the coherent 
state with the $\beta$ correction can be found. Following Appendix A the formula can be written as
\begin{align}
r = \frac{\beta}{\hslash} \sqrt{X^2+Y^2}+\mathcal{O}(\beta^2),
\end{align}
where the $X$ and $Y$ functions can be read out from the 
expressions \eqref{CoherentSecondMoments1}--\eqref{CoherentSecondMoments3}.

Similar to the study of the first-order moments, the phase $\theta$ is of no importance for our discussion and is therefore fixed at $\theta=0$ for the simplicity of the 
analysis. Introducing $\phi :=\omega t$, it follows:
\begin{widetext}
\begin{align}
\frac{r}{\beta \omega \hslash} &= 
\frac{1}{6}  \left[ 4 \sin ^2(\phi ) \left[|\alpha|^2 \left(2 (|\alpha|-1) |\alpha|^2 \sin (3 \phi )+3 \left(2 |\alpha|^2-3\right) \phi  \cos (\phi )+(9 |\alpha| -3) \sin (\phi )\right)+3 \sin (\phi)\right]^2 \right. \nonumber \\
&+ \left. \left[\sin (2 \phi ) \left(3 |\alpha|^4+8 |\alpha|^2+4 |\alpha| ^4 \cos (2 \phi )+3\right)-6 |\alpha| ^2 \left(2 |\alpha| ^2+3\right) \phi  \cos (2 \phi )\right]^2 \right
]^{1/2} +\mathcal{O} \left( \beta \right).
\label{rbetaomegah}
\end{align}
\end{widetext}
Importantly, this expression exhibits a factor linear 
in $\phi$, resulting in the accumulation of squeezing over 
time. This is in contrast to the case of energy eigenstates $\ket{n}$
discussed before. At late times, i.e., $\phi \gg 1$, the term 
with the contribution linear in $\phi$ dominates over the 
accompanying oscillatory ones and the formula 
(\ref{rbetaomegah}) is well approximated by
\begin{widetext}
\begin{align}
\frac{r}{\beta \omega \hslash} &\approx \frac{\phi |\alpha|^2}{2 \sqrt{2}} \sqrt{45+36 |\alpha|^2+20 |\alpha|^4+3 \left(4 |\alpha|^4+20 |\alpha| ^2+9\right) \cos (4 \phi)}.
\label{rbetaomegahapprox}
\end{align}
\end{widetext}
In consequence, at late times, the squeezing exhibits oscillatory behavior with frequency 
$4\omega$, and the amplitude of the oscillations belonging to the range $[r_{\text{min}},r_{\text{max}}]$,
where
\begin{align}
r_{\text{min},\ket{\alpha}} &\approx  \beta \omega \hslash \frac{1}{2} \left|\left(2 |\alpha| ^2-3\right)\right| |\alpha| ^2 \omega t
\label{rbetaomegahmin}
\end{align}
and 
\begin{align}
r_{\text{max},\ket{\alpha}} &\approx  \beta \omega \hslash \left|\left(2 |\alpha| ^2+3\right)\right| |\alpha| ^2 \omega t.
\label{rbetaomegahmax}
\end{align}
This formula suggests that the squeezing can grow 
to a relevant magnitude after a sufficiently long 
time. The possible associated empirical consequences will be examined in Sec. \ref{Sec:experiments}. Care should, however, be taken here as this cumulative effect, when integrated over long times, leads to a breakdown of the perturbative approach used in this study. This happens for times typically greater than $\left( \beta \omega^2 \hslash \right)^{-1}$. Naturally, the lower the photon energy is compared to the energy scale of new physics $\beta^{-1}$, the longer the perturbative approach remains valid. The extension of this work to the nonperturbative regime is beyond the scope of this paper and left for future work. 

\begin{figure}[h!]
    \centering
    \includegraphics[scale=0.55]{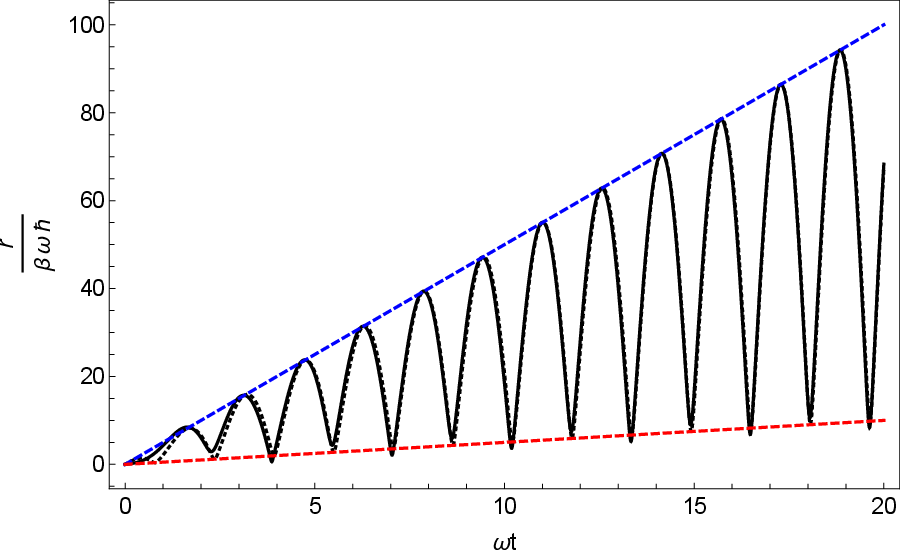}
    \caption{Time evolution of the squeezing amplitude $r$ for $|\alpha|=1$. 
    Here, the solid (black) line corresponds to the formula given by Eq. 
    \eqref{rbetaomegah}. The approximation provided in Eq. \eqref{rbetaomegahapprox} 
    is depicted as a dotted (black) line. The upper dashed (blue) line 
    corresponds to Eq. \eqref{rbetaomegahmax}, and the bottom dashed (red)
    line corresponds to Eq. \eqref{rbetaomegahmin}.
    }
    \label{fig:SqeezAmp}
\end{figure}

In Fig. \ref{fig:SqeezAmp}, we demonstrate 
the time dependence given by the formula (\ref{rbetaomegah}) 
alongside its approximation (\ref{rbetaomegahapprox}).
The approximation proves to be highly effective, 
exhibiting only minor deviations from the original formula 
at short timescales. Additionally, Fig. \ref{fig:SqeezAmp} 
includes the approximations for $r_{\text{min},\ket{\alpha}}$
and $r_{\text{max},\ket{\alpha}}$, as described by Eqs.
(\ref{rbetaomegahmin}) and (\ref{rbetaomegahmax}), respectively. 
These approximations also maintain a high degree of accuracy, 
closely fulfilling their intended roles.

\section{Deformed dispersion relation}
\label{Sec:dispersion}

Most of the relevant properties of light (except for the 
polarization states) can be captured by considering a 
massless scalar field model.  This allows one to 
consider the multimode light state and derive the 
associated dispersion relation. 

The multimode light Hamiltonian generalizes Eq. \eqref{SingleModeH} to
\begin{align}
\hat{H} &= \sum_{{\bf k}} \hat{H}_{\bf k} = \frac{1}{2} \sum_{{\bf k}} \left( \hat{P}^2_{\bf k} +\omega^2 \hat{Q}^2_{\bf k}   \right) \nonumber \\
            &= \frac{1}{2} \sum_{{\bf k}} \left( \hat{p}^2_{\bf k} +\omega^2 \hat{q}^2_{\bf k}   \right)+ \frac{\beta}{3} \sum_{{\bf k}}\hat{p}^4_{\bf k}+\mathcal{O}(\beta^2), 
\end{align}
where  $\omega^2= c^2 {\bf k}  \cdot {\bf k} $, which is just a sum of 
single-mode Hamiltonians for different values of $\omega$. 
As a consequence, previous results, and in particular the energy levels and 
perturbed eigenstates, can be adopted here. 

The $\hat{q}_{\bf k}$ and $\hat{p}_{\bf k'}$ satisfy the 
canonical commutation relation $[\hat{q}_{\bf k},\hat{p}_{\bf k}]=
i \hslash \hat{\mathbb{I}}\delta_{{\bf k},{\bf k}}$.

Employing the expression for the first-order vacuum state 
$\ket{0^{(1)}}$ given by Eq. \eqref{01state}, one can evaluate 
the action of the field operator on that state:
\begin{equation}
\hat{q}_{\bf k}(0) \ket{ 0^{(1)}} = \sum_n c_n \ket{n^{(1)}} \, . 
\end{equation}
At first order in $\beta$,

\begin{align}
c_1 &=\sqrt{\frac{\hslash}{2\omega}}\left(1+\frac{\beta \hslash\omega}{2}\right) \,,\\
c_3 &= - \frac{\sqrt{3}}{4} \beta  \hslash^{3/2} \omega^{1/2} \,,
\end{align}
and the remaining coefficients $c_n$ are equal to zero.
Single-particle states are therefore not the only states created by an elementary excitation of the vacuum: three-particle states are also expected.  While Eq. \eqref{01state} may suggest that 
also five-particle states are created, this contribution cancels out at first order in $\beta$. This property has further consequences on the propagation of the quanta. 

To quantify this effect, the 
two-point correlation function of $\hat{q}(\Vec{x},t)$ on the 
first-order vacuum state $\ket{0} := \bigotimes_{\bf k} \ket{0^{(1)}_{\bf k}} 
\in \mathcal{H}$ is considered:
\begin{eqnarray}
\langle 0| \hat{q}({\bf x},t) \hat{q}({\bf y},t') |0 \rangle &=& 
\frac{1}{V} \sum_{{\bf k},n} |c_n|^2 e^{i {\bf k \cdot (x - y)} - i \frac{\Delta E_n^{(1)}}{\hslash} (t - t')} \nonumber \\
&=& \frac{1}{V} \sum_{{\bf k}} \int \frac{d\omega}{2\pi} G_{p} e^{i {\bf k \cdot (x - y)} - i \omega (t - t')} , \nonumber \\ 
\label{two point function correlation}
\end{eqnarray}
where 
\begin{equation}
\Delta E_n^{(1)} = E^{(1)}_{n} - E^{(1)}_{0} = \hslash \omega n
\left(1+\frac{\beta \hslash \omega}{2}\left(n + 1\right)\right) \, .
\end{equation}

The $G_{p}$ entering Eq. (\ref{two point function correlation}) is a propagator. Introducing the four-momentum $p=(\frac{\hslash \omega}{c}, 
\hslash {\bf k})$, of pseudonorm $p^2 = -\frac{\hslash^2\omega^2}{c^2} + \hslash^2 {\bf k}  \cdot {\bf k}$, it writes
\begin{align}
G_{p} :=& \sum_{n} \frac{2i \Delta E_n^{(1)} |c_n|^2}{p^2 + (\frac{\Delta E_n^{(1)}}{c})^2 - {\bf k}  \cdot {\bf k} - i\epsilon} \nonumber \\
=& \frac{i\left(\hslash^2 +2\beta \hslash^3 \omega\right)}{
p^2+\frac{2\beta\hslash^3\omega^3}{c^2}- i\epsilon}+\mathcal{O}(\beta^2) \, .
\label{Propagator}
\end{align}
 
Because of to the $c_n$ coefficients appearing with a modulus squared, only $n=1$ contributes to the previous expression at first order in $\beta$, the $n=3$ term being of the order $\beta^2$. In the $\beta\rightarrow 0$
limit the standard Feynman propagator of a scalar field is 
recovered: $G_{p}=\frac{i}{p^2- i\epsilon}$.  

As we see, the propagator (\ref{Propagator}) exhibits a pole at
\begin{equation}
p^2 c^2+2\beta\hslash^3 \omega^3=0 \, ,
\end{equation}
which corresponds to the dispersion relation
\begin{equation}
\omega = c k+\beta \hslash c^2 k^2+\mathcal{O}(\beta^2) \, , 
\end{equation}
where $k$ denotes the norm of the wave vector ${\bf k}$. 
This expression exhibits an additional term $\propto k^2$. 
As expected, the usual linear form is recovered in the 
$\beta \rightarrow 0$ limit.

This quantum correction to the dispersion relation 
leads to an energy dependence of the photon group velocity:
\begin{equation}
v_{\text{gr}} : =\frac{\partial \omega}{\partial k} = c +2\beta \hslash c^2 k \, . 
\end{equation}

For positive values of $\beta$, the higher the photon energy, 
the faster it moves. This cumulative effect leads to a 
time advancement as the photon energy increases. For two photons 
$1$ and $2$ of energies $E_{1}$ and $E_{2}$ the difference in
time arrival $\Delta t := t_{2} - t_{1}$ writes
\begin{equation}
c \Delta t \approx - 2 \beta \Delta E L,
\label{Delta}
\end{equation}
$L$ being the traveled distance and $\Delta E := E_{2} - E_{1}$. 
The correct limit $\Delta t \rightarrow 0$ when $\beta \Delta E \rightarrow 0$
is recovered. Please note that the formula (\ref{Delta}) does not 
take into account any effect of cosmological expansion.

\section{Observability} 
\label{Sec:experiments}

Testing quantum gravity effects using astrophysical photons 
is an intensively explored avenue in quantum gravity phenomenology \cite{Ellis:2011ek, FermiGRB, Amelino-Camelia:2016ohi, Barrau:2016fcg, Addazi:2021xuf, Ng:2022sjc, He:2022gyk}. By 
studying the behavior of photons, particularly those originating 
from distant astrophysical sources such as gamma ray bursts, 
one can attempt to probe the fundamental nature of gravity at the 
quantum level \cite{Amelino-Camelia:1997ieq, Ellis:2002in, Amelino-Camelia:2017zva, Bernardini:2017tzu, Bartlett:2021olb}. These experiments 
involve examining subtle deviations from classical predictions, 
such as the energy-dependent speed of photons and the resulting 
time lags or advancement accumulated over long distances. An 
example of such an accumulative effect has been provided in 
the previous section for the GUP-type effects. 

To estimate the magnitudes of both the squeezing and time advancement of photons, it has to be remembered that the parameter $\beta$ has the dimension $\left[ \beta \right] = \left[ E^{-1} \right]$. Since the considered modification to the usual uncertainty principle is expected to be 
a manifestation of Planck-scale physics, $\beta^{-1}$ is expected to be around the Planck scale:
\begin{equation}
\beta \sim \frac{1}{E_{\text{Pl}}},
\end{equation}
where $E_{\text{Pl}} \approx 1.22 \times 10^{19}$ GeV.\\

The time-advancement established equation (\ref{Delta}) 
can be written as
\begin{equation}
\Delta t \sim  -\left(\frac{\Delta E}{E_{\text{Pl}}}\right) \frac{L}{c} .
\label{DeltaEpl}
\end{equation}

Two high-energy photons in the energy range of the Cherenkov Telescope Array (CTA) \cite{CTAConsortium:2017dvg}, typically of 1 and 10 TeV, respectively,\footnote{The full energy range covered by CTA being comprised between 20 GeV and 300 TeV.} would undergo a time advancement of $|\Delta t| \sim 7$ s after a traveled distance of 100 Mpc.\\

This article aimed to explore the potential cumulative effects on 
phase-space properties of quantum states, specifically squeezing.
As shown in Sec. \ref{Sec:squeezing}, cumulative effects are not expected for energy eigenstates at linear order in the $\beta$ parameter. They are, however, predicted for coherent states with $\alpha\neq 0$. 

Indeed, the expected maximal squeezing of the vacuum state $\ket{0}$ is,  according to Eq. \eqref{rmaxvacuum},

\begin{equation}
r_{\text{max}, \ket{0}} \sim \frac{E}{E_{\text{Pl}}},
\end{equation}
which, for $E \sim 1$ TeV photons, gives a very small value $r_{\ket{0}}  \sim 10^{-16}$.

On the other hand, for coherent states, the formula 
(\ref{rbetaomegahmax}) can be recast into 
\begin{equation}
r_{\text{max}, \ket{\alpha}} \sim \left|\left(2 |\alpha| ^2+3\right)\right||\alpha| ^2 \left(\frac{E}{E_{\text{Pl}}}\right)^2 \frac{L}{l_{\text{Pl}}}, 
\end{equation}
which for $|\alpha| \sim 1$ reduces to the following estimate: 
\begin{equation}
r_{\text{max}, \ket{\alpha}} \sim \left(\frac{E}{E_{\text{Pl}}}\right)^2 \frac{L}{l_{\text{Pl}}}.
\label{squeezing max numerical evaluation}
\end{equation}

The validity of the perturbative approach used in this work requires the previous estimation of $r_{\text{max}}$ to be smaller than unity. For photons of fixed energy, this imposes a maximum distance of $l_{\text{Pl}} (E_{\text{Pl}}/E)^2$. For 1 eV photons, this distance is around $10^5$ parsecs, and its value quickly decreases as the photon energy goes up. For higher-energy photons at this distance, although the perturbative expansion cannot be used anymore, important squeezing effects are still to be expected.

The important takeaway message is, therefore, that photons of astrophysical origin emitted in a coherent state should undergo important squeezing due to the GUP, although the precise computation of the value of the effect for high-energy photons requires one to go beyond the perturbative approach.\\

\emph{Homodyne detectors} 
(see e.g. \cite{Lvovsky:2014sxa}) offer a promising means to measure squeezing in quantum systems. These detectors 
are capable of detecting both the amplitude and phase information 
of a quantum state, making them well-suited for investigating 
the squeezing phenomenon. By employing homodyne detection techniques, 
one can measure the variances of both quadrature components of the
quantum state, which directly relates to the degree of squeezing
present. Employing this technique in analyzing photons of 
astrophysical origin could serve as a new window of constraining 
quantum gravity effects. 

Furthermore, analysis of the statistics of light may also provide 
an opportunity to constrain quantum gravitational effects. 
In particular, the projection in the interacting theory of the coherent states onto the occupation number ones [in the leading order -- $\mathcal{O}(\beta)$] leads to the following probabilities:
\begin{align}
    P_n :=& |\langle n^{(1)} | \alpha(t) \rangle |^2 \\
    =& \frac{|\alpha|^{2n}}{n!} e^{-|\alpha|^2}+\frac{\beta\hslash\omega}{12} e^{-|\alpha|^2}  \nonumber \\
    &\times \left[- \frac{1}{2} \frac{|\alpha|^{2n+4}}{n!} \cos\left(4\theta\right) + \frac{4n+6}{n!} |\alpha|^{2n+2} \cos\left(2\theta\right) \right. \nonumber\\
    &-\left.  \frac{4n-2}{\left(n-2\right)!} |\alpha|^{2n-2} \cos\left(2\theta\right) + \frac{1}{2} \frac{|\alpha|^{2n-4}}{\left(n-4\right)!} \cos\left(4\theta\right) \right]\,, \nonumber
\end{align}
exhibiting a correction that is linear in $\beta$. No accumulation 
of the correction is, however, expected at this order.

\section{Summary}
\label{Sec:summary}

This article explored potential effects of Planck-scale physics 
on the Heisenberg uncertainty principle. The focus has been put on the time evolution 
of quantum states of single-mode light under the influence of quantum 
gravitational corrections.\\ 

Findings show that for both energy eigenstates and coherent states of light, 
the leading order quantum gravity corrections result in squeezing following 
an oscillatory pattern. There is, however, a significant difference 
between the two cases.

For the energy eigenstates, there is no net cumulative effect at first order 
in the parameter $\beta$ governing the strength of Planck-scale effects. 
Those states remain stable under the influence of the proposed quantum gravity 
effects, and no significant contribution from the Planck scale is expected 
in this case.

For the coherent states such that $\alpha \neq 0$, there is, however, a net accumulation 
of the amplitude of squeezing (and displacement) over time. The effect is predicted 
to be strong in the case of photons traveling on astrophysical or cosmological distances 
and may have potential empirical constraints. For high-energy photons, the predictions require going beyond the linear-perturbation approach used here, which will be the object of a future paper. Measurements of nonclassical properties 
of light originating from distant astrophysical sources may, therefore, open a window 
to test Planck-scale physics through such predictions. 

A corrected form of the dispersion relation of light has also been derived that 
provides additional insights, including the analysis of corresponding advancements 
of photons. Interestingly, both the constraints on the time of arrival and the squeezing 
can be used simultaneously, hopefully leading to tighter bounds on the effects. 
This possibility opens another interesting path for further studies.

Future research in this direction may also expand the investigation
upon considering the generalized extended uncertainty principle \cite{Hossenfelder:2012jw}: 
\begin{equation}
\Delta Q \Delta P \geq \frac{\hslash}{2}  (1+\beta \Delta P^2+\alpha \Delta Q^2). \label{GUP:QP}
\end{equation}
This generalized form of the uncertainty principle introduces 
additional terms $\beta$ and $\alpha$, which modify the 
trade-off between the uncertainties in position ($\Delta Q$) 
and momentum ($\Delta P$). To go even beyond, one could also include linear terms in ($\Delta P$) and/or in ($\Delta Q$), in the spirit of \cite{Das:2010zf,Vagenas:2017vsw}. An interesting objective is also 
to investigate the fate of the thermal states, not considered here,
and the effect on the polarization degrees of freedom of light. 

\section{Acknowledgments}

J.M. was supported by Grants No. DEC-2017/26/E/ST2/00763 
and No. UMO2018/30/Q/ST9/00795 from the National Science Centre, 
Poland.

\section*{Appendix A}

Let us consider the following form of the variances and of the covariance of the operators $\hat{q}$ and $\hat{p}$: 
\begin{align}
\omega (\Delta \hat{q})^2 &= \frac{\hslash}{2}+X\beta +A \beta^2 +\mathcal{O}(\beta^3), \\
 (\Delta \hat{p})^2/\omega &= \frac{\hslash}{2}-X\beta +B \beta^2 +\mathcal{O}(\beta^3), \\
C_{qp} &= Y \beta + C \beta^2 +\mathcal{O}(\beta^3),
\end{align}
where $X,Y,A,B,C$ are some functions, not being dependent on $\beta$. 
In this article, the forms of $X$ and $Y$ are derived, while 
the form of the $A, B, C$ functions is not known. 

By applying the above expressions to Eq. (\ref{eigenvaluesLambda})
one finds that
\begin{align}
\lambda_{\pm} = \frac{\hslash}{2}\pm \sqrt{X^2+Y^2} \beta +\mathcal{O}(\beta^2).
\end{align}
Therefore, the $\mathcal{O}(\beta^2)$ terms, related to 
the unknown functions $A, B,$ and $C$, entering the square 
root in Eq. \eqref{eigenvaluesLambda}, cancel out and do not 
contribute in the linear order in $\beta$ to the eigenvalues 
$\lambda_{\pm}$.

Consequently, by expressing the square roots of the 
eigenvalues as follows:
\begin{align}
\sqrt{\lambda_{\pm}} = \sqrt{\frac{\hslash}{2}} e^{\pm r},
\end{align}
we find that the amplitude of squeezing can be written as
\begin{align}
r = \frac{\beta}{\hslash} \sqrt{X^2+Y^2}+\mathcal{O}(\beta^2).
\end{align}
This formula is used to derive the expressions for the 
amplitude of squeezing in Sec. \ref{Sec:squeezing}.

 \end{document}